\documentclass[letterpaper, 10pt, conference]{ieeeconf}
\IEEEoverridecommandlockouts                              
\usepackage{amsmath}

\usepackage{xcolor}
\usepackage[utf8]{inputenc} 
\usepackage[T1]{fontenc}
\usepackage{url}
\usepackage{amsfonts}                                                          
\usepackage{epsfig}

\usepackage{graphicx}        
\usepackage{latexsym}
\usepackage{amssymb}
\usepackage{multirow}
\usepackage{cite}
\usepackage{mathtools}
\usepackage{epstopdf}
\usepackage{etoolbox}
\usepackage{array}
\usepackage{mathptmx}
\usepackage[bottom]{footmisc}
\usepackage{tikz}

\usetikzlibrary{decorations.pathreplacing}

\usepackage{verbatim}
\usepackage{calrsfs}
\usepackage{booktabs}
\newcommand{\ES}[1]{\textcolor{black}{{#1}}}

\newcommand{\CDC}[1]{\textcolor{black}{{#1}}}

\newcommand\numeq[1]%
  {\stackrel{\scriptscriptstyle(\mkern-1.5mu#1\mkern-1.5mu)}{=}}

\DeclareMathAlphabet{\pazocal}{OMS}{zplm}{m}{n}

\DeclareMathOperator{\trace}{tr}
\DeclareMathOperator{\diag}{diag}
\newcommand{\Cov}{\mathrm{cov}}
\let\bbordermatrix\bordermatrix
\patchcmd{\bbordermatrix}{8.75}{4.75}{}{}
\renewcommand{\theequation}{\arabic{equation}}
\renewcommand{\thefigure}{\arabic{figure}}
\renewcommand{\thetable}{\arabic{table}}

\newcommand{\T}{^{\mbox{\tiny T}}}

\newcommand{\rar}{\rightarrow}

\newcommand{\tri}{\triangleq}

\newcommand{\be}{\begin{equation}}
\newcommand{\ee}{\end{equation}}
\newcommand{\bea}{\begin{eqnarray}}
\newcommand{\eea}{\end{eqnarray}}
\newcommand{\bes}{\begin{eqnarray*}}
\newcommand{\ees}{\end{eqnarray*}}
\newcommand{\bce}{\begin{center}}
\newcommand{\ece}{\end{center}}

\newcommand{\eeae}{\end{IEEEeqnarray}}

\def\VR{\kern-\arraycolsep\strut\vrule &\kern-\arraycolsep}
\def\vr{\kern-\arraycolsep & \kern-\arraycolsep}

\newcommand{\ben}{\begin{enumerate}}
\newcommand{\een}{\end{enumerate}}

\newcommand{\hso}{\hspace{.1in}}
\newcommand{\hst}{\hspace{.2in}}

\newtheorem{theorem}{Theorem}

\newtheorem{remark}{Remark}
\newtheorem{corollary}{Corollary}

\newtheorem{lemma}{Lemma}
\newtheorem{example}{Example}

\newtheorem{proposition}{Proposition}

\usepackage[utf8]{inputenc} 
\usepackage[T1]{fontenc}
\usepackage{url}
\usepackage{ifthen}
\usepackage{cite}

\title{Indirect Rate Distortion Functions with Side Information:    Structural Properties  and  Multivariate  Gaussian  Sources}

\author{Evagoras Stylianou,  Michail Gkagkos and  Charalambos D. Charalambous
\thanks{E. Stylianou is with the Chair of Theoretical Information Technology, Technical University of Munich, Munich, Germany %
        {\tt\small evagoras.stylianou@tum.de}}%
\thanks{M. Gkagkos is with the Department of Electrical and Computer Engineering, Texas A\&M University, College Station, Texas, USA
        {\tt\small gkagkos@tamu.edu}}%
\thanks{C. D. Charalambous is with the Department of Electrical and Computer Engineering, University of Cyprus, Nicosia, Cyprus
        {\tt\small chadcha@ucy.ac.cy}}%
}

\begin{document}


\maketitle

\begin{abstract}
\ES{In this paper, we analyze the indirect source coding problem with side information at both the encoder and decoder, as well as only at the decoder. We first derive structural properties of the two rate distortion functions (RDFs) for general abstract spaces and identify conditions under which the RDFs coincide. For multivariate jointly Gaussian random variables with square-error fidelity, we establish structural properties of the optimal test channels, show that side information at both the encoder and decoder does not reduce compression, and provide water-filling solutions using parallel Gaussian channel realizations.} \CDC{ This paper uses  a novel realization theory approach to establish   achievability of the converse coding theorem  lower bounds of the two  RDFs.}

\end{abstract} \vspace{-0.1cm}
\section{Introduction}
\label{sect:problem}

\ES{Shannon's rate distortion function (RDF) \cite{shannon1959coding} characterizes the minimum rate required to compress a source while satisfying a fidelity criterion,} \CDC{i.e.,    the optimal performance theoretically attainable by noncausal codes.} \ES{This is known as the \emph{source coding problem} (see also \cite{Gallager:1968, berger:1971}). Kolmogorov's $\epsilon$-entropy \cite{kolmogorov:1959} extends the RDF from finite spaces to abstract spaces like Borel spaces. Both the RDF and $\epsilon$-entropy are essential mathematical tools for analyzing communication sources and systems.}

\begin{figure}[t]
\centering
  \includegraphics[width = \columnwidth]{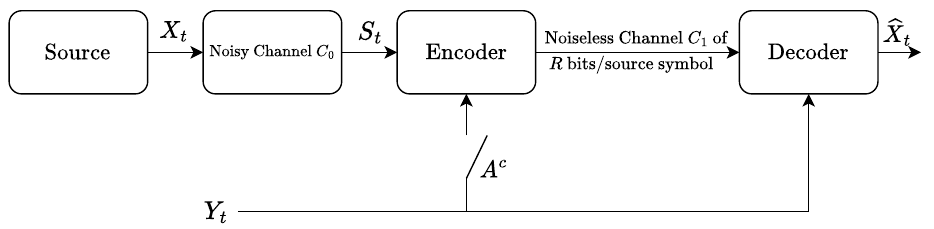}\vspace{-0.2cm}
 \caption{When switch $A$ is closed, side information is available at both the encoder and decoder; when open, it is available only at the decoder.
} \vspace{-0.65cm}
  \label{fg:blockdiagram} 
\end{figure}

\ES{Dobrushin and Tsybakov \cite{dobrushin1962information} introduced the indirect source coding problem, where the encoder observes the source through a noisy channel.} \CDC{Witsenhausen \cite{witsenhausen1980indirect} revisited \cite{dobrushin1962information} }\ES{ for finite alphabet random variables (RVs) with Hamming distance distortion. Yamamoto et al. \cite{yamamoto1980source} and Draper and Wornell \cite{draper-wornell2004} computed the RDF for  the indirect source coding problem with side information only at the decoder for scalar Gaussian RVs, corresponding to an open switch \(A\) in Figure~\ref{fg:blockdiagram}. It is noted that by replacing the noisy channel \(C_0\) with a noiseless one reduces the problem to Wyner-Ziv \cite{wyner-ziv1976} or Wyner \cite{wyner1978}, and without side information, it becomes the classical source coding problem \cite{Cover-2006}. Tian and Chen \cite{tian-chen2009} and Zahedi et al. \cite{Zahedi-Ostegraard-2014} extended} \CDC{the indirect source coding problem} \ES{to multivariate Gaussian RVs, but as discussed in \cite{gkagkosISIT, gkagkos2024structural}, their solutions do not match the multivariate Wyner RDF \cite{wyner1978} when \(C_0\) is noiseless.} \CDC{In this paper, we consider  the indirect source coding problem for multivariate Gaussian RVs, and }\ES{we present alternative closed-form solutions  that correctly align with Wyner's \cite{wyner1978, gkagkos2024structural} and the classical RDFs~\cite{Cover-2006}.}

\ES{The indirect source coding problem has broad applications, including multiterminal and distributed lossy compression problems \cite{Oohama2005, Oohama1997, ViswanathanCEO1997, SUlukus2012, JunChen2014, Cheng2004}. It is also relevant in distributed filtering, such as extending Bucy's \cite{bucy:1982} mean-square error filtering, and in nonanticipative RDFs \cite{stavrou-charalambous-charalambous-loyka2018siam}. Additionally, it applies to designing encoders and decoders for Gaussian channels with memory \cite{charalambous-kourtellaris-tziortzis:SICON-2024, ISIT:NonanticipativeRDF}, where the side information \(Y^n = \{Y_t \mid t=1, \ldots, n\}\) is generated by another noisy channel with input \(X^n\), and \(\widehat{X}^n\) estimates \(X^n\) under a fidelity constraint. Distributed estimation applications are discussed in Remark~\ref{rem:appl_f}.}

 \ES{This paper examines the indirect source coding problem in Figure~\ref{fg:blockdiagram}. We derive structural properties for the optimization problems of the RDFs for scenarios with side information at both the encoder and decoder, as well as, at the decoder alone, for abstract spaces. We also identify conditions under which the two RDFs are equal.} \CDC{Subsequently, we calculate the RDFs for multivariate Gaussian sources, channel \(C_0\), and side information, all under square-error distortion. Using realization theory, we construct realizations of \(\widehat{X}^n\) that achieve the RDFs and ensure the converse coding lower bound is met. Finally, we derive water-filling solutions for the RDFs.}
 


 
 \subsection{Problem Statement and Discussion}
 The indirect source coding problem  of Figure~\ref{fg:blockdiagram},  is defined as follows. On a given probability space $(\Omega, \pazocal{F}, {\mathbb P})$  we consider   a triple of jointly independent and identically distributed (IID) RVs, 
\begin{align*}
&(X^n, S^n, Y^n)= \{ (X_t, S_t,Y_t): t=1, 2, \ldots, n\},  \\
& X_t : \Omega \rar {\mathbb X},\hso S_t : \Omega \rar {\mathbb S}, \hso Y_t : \Omega \rar {\mathbb Y}, \\
 &{\bf P}_{X^n,S^n,Y^n} = \times_{t=1}^n {\bf P}_{X_t,S_t,Y_t}, \: {\bf P}_{X_t,S_t,Y_t}={\bf P}_{X,S,Y}, \: t=1, \ldots, n , 
 \end{align*}
where $({\mathbb X}, {\cal B}({\mathbb X})),  ({\mathbb S}, {\cal B}({\mathbb S})), ({\mathbb Y}, ({\cal B}({\mathbb Y}))$ are standard   Borel spaces  (i.e., complete, separable metric spaces). We  wish to reconstruct realizations of \CDC{the RV,  $X^n=x^n$, by  realizations of the RV,    $\widehat{X}^n=\widehat{x}^n$, }    $\widehat{X}_t : \Omega \rar \widehat{\mathbb X}\subseteq {\mathbb X}$,  with respect to the average fidelity or   distortion fuction,     
\begin{align*}
\frac{1}{n}{\bf E} \Big ( \sum_{t=1}^n d(X_t,\widehat{X}_t)\Big ) \leq \Delta, \hso d: {\mathbb X} \times \widehat{\mathbb X} \rar [0,\infty),
 \end{align*}
where $d(\cdot, \cdot)$ is a measurable function.
 
The operational definition of an achievable compression rate, i.e., the encoder, decoder, etc., is a slight variation of the one given in \cite{wyner-ziv1976,wyner1978} (see also \cite{dobrushin1962information,Gallager:1968,berger:1971}). \ES{We analyze two scenarios: side information at both the encoder and decoder, and side information only at the decoder.}

 {\it Switch $A$ Open ($A^\mathrm{c}$).} 
\ES{When Switch $A$ is open, side information $\{Y_t: t=1,2,\ldots,n\}$ is available only at the decoder. According to Wyner \cite{wyner1978} and Yamamoto et al. \cite{yamamoto1980source}, the infimum of all achievable rates for average distortion $\Delta \in [0,\infty)$, is characterized by the single-letter conditional RDF:}
\begin{align}
{R}_{S;Z|Y}(\Delta) \tri& \inf_{ \pazocal{D}_{A^\mathrm{c}}(\Delta)} \big\{I(S;Z)-I(Y;Z)\big\}, \nonumber \\ = &  \inf_{\pazocal{D}_{A^\mathrm{c}}(\Delta)} I(S; Z|Y), \label{rdf_d1}
\end{align}
where  $\pazocal{D}_{A^\mathrm{c}}(\Delta)$ is specified by the set of auxiliary RVs $Z$,
\begin{align}
\pazocal{D}_{A^\mathrm{c}}(&\Delta)\tri \big\{ Z: \Omega \rar {\mathbb Z} \big | \; {\bf P}_{Z|X,Y,S}={\bf P}_{Z|S}, \;\exists \: \mbox{measurable funct.} \nonumber \\ & f: {\mathbb Y}\times {\mathbb Z} \rar \widehat{\mathbb X}, \; \widehat{X}=f(Y,Z), \;{\bf E}\big(d(X, \widehat{X})\big)\leq \Delta \big\}, \label{con_1}
\end{align}
and the joint  measure  ${\bf P}_{X, S, Y, Z, \widehat{X}}$ induced by the RVs $(X,S, Y, Z, \widehat{X})$ utilizes  the   ``test channel'' ${\bf P}_{Z|S}$ from $S$ to $Z$. {At this point we note that  ${R}_{S;Z|Y}(\Delta)$ is a decentralized optimization problem with multiple strategies (as indicated by the set (\ref{con_1})), which do not share the same information.}

 \ES{\noindent {\it Switch $A$ Closed ($A$).} When Switch $A$ is closed, side information}  \CDC{ $\{Y_t: t=1,2,\ldots,n\}$} \ES{ is available at both the encoder and the decoder.} \CDC{By slight variation of } \ES{Wyner \cite{wyner1978}, the infimum of all achievable rates, denoted by $R_{A}(\Delta)$, for average distortion $\Delta \in [0,\infty)$, is given by the single-letter conditional RDF:}
\begin{align}
{R}_{S;\widehat{X}|Y}(\Delta) \tri& \inf_{\pazocal{D}_{A}(\Delta)}  I(S; \widehat{X}|Y), \hso \Delta \in [0,\infty), \label{eq:OP1ck1}
\end{align}
where $\pazocal{D}_{A}(\Delta)$ is specified by the set 
$$
\pazocal{D}_{A}(\Delta)\triangleq \big\{ \widehat{X}: \Omega \rar \widehat{\mathbb X} \big |  {\bf P}_{\widehat{X}|S,Y, X}={\bf P}_{\widehat{X}|S,Y}, \ \ {\bf E}\big(d(X, \widehat{X})\big)\leq \Delta \big\}$$
and the  joint  measure  ${\bf P}_{X, S, Y,  \widehat{X}}= {\bf P}_{\widehat{X}|S,Y}{\bf P}_{S,Y,X}$ induced by the RVs $(X,S, Y,  \widehat{X})$, utilizes the test channel ${\bf P}_{\widehat{X}|S,Y}$. 


\begin{remark}\label{rk:class}
 We note that if $X = S-$a.s. then the RDFs ${R}_{S;Z|Y}(\Delta)$, ${R}_{S;\widehat{X}|Y}(\Delta)$ reduced to Wyner's RDFs \cite{wyner1978}.
If in addition, $Y$ is nonrandom
then both RDFs are equal to the classical RDF \cite{Cover-2006}.\vspace{-0.35cm}
\end{remark}
\CDC{\begin{remark} 
    Papers \cite{gkagkos2024structural,gkagkosISIT}, showed that the RDF $ {R}_{S;Z|Y}(\Delta)$ from \cite[Theorem~4]{tian-chen2009} is incorrect and identified gaps in the derivation. Specifically,  the solution of \cite[Theorem~4]{tian-chen2009} does not reproduce Wyner's RDF \cite{wyner1978} when $X=S$-a.s..  (and  similarly for the solution of  \cite[Theorem~3]{Zahedi-Ostegraard-2014}).
\end{remark}}

 \subsection{Notation} 
\label{sec:notation}
\CDC{Let  ${\mathbb Z}$,  ${\mathbb Z}_+$, $\mathbb{R}$,}  be the set of integers, positive integers, {and} real numbers,  respectively. The expression $\mathbb{R}^{n \times m}$ {for $n,m \in {\mathbb Z}_+$,  denotes the set of $n$ by $m$ matrices with entries the real numbers. Let $||x||$ \CDC{denote the Euclidean  norm of} $x \in  \mathbb{R}^{n}$. For a symmetric matrix $Q \in {\mathbb R}^{n \times n}$, the notation  $Q \succeq 0$ (resp. $Q \succ 0$) means  the matrix is positive semi-definite (resp. positive definite). The notation $Q_2 \succeq Q_1$ means that $Q_2 - Q_1 \succeq  0$. For any  matrix $A\in \mathbb{R}^{p\times m}$, we denote its transpose by $A\T$,  and for $m=p$,  we denote its trace and its determinant  by  $\trace(A)$ and $\det\big(A\big)$, respectively. By $\diag(A)$ we denote the diagonal matrix with entries those of $A$ and zero elsewhere. By  $I_p$ we denote the ${p\times p}$ identity matrix. For RVs $X$, $Y$, and $Z$, $I(X;Y|Z)$ denotes the conditional mutual information given $Z$ \cite{Gallager:1968}.  Any  triple of jointly Gaussian RVs  $X: \Omega \rar {\mathbb R}^{n_x} , S:\Omega \rar {\mathbb R}^{n_s}, Y:\Omega \rar {\mathbb R}^{n_y}$ will be denoted by $(X,S,Y)\in G(0, Q_{(X,S,Y)})$.
The conditional covariance of \CDC{arbitrary} RVs $X$ and $Y$ conditioned on the \CDC{arbitrary} RV $Z$ is denoted by $Q_{X,Y|Z} \tri \Cov(X,Y|Z) $ where,\vspace{-0.1cm}
\begin{align*}
\Cov(X,Y|Z) &\tri  {\bf E} \big((X- {\bf E}(X\big|Z))(Y- {\bf E}(Y|Z))\T \big|Z   \big),\\
& \numeq{1} {\bf E} \big ((X- {\bf E}(X|Z))(Y- {\bf E}(Y|Z))\T    \big ),
\end{align*}
where (1) holds if $(X, Y, Z) \in  G(0, Q_{(X,Y,Z)})$.
Similarly, the conditional covariance of RV $X$ given RV $Z$ is denoted by $Q_{X|Z} \tri \Cov(X,X|Z) $.

\subsection{Main Contribution and Organisation}
\ES{In this section, we present our main result: the water-filling solution for the two RDFs. For jointly Gaussian RVs \((X, S, Y)\) with a square-error distortion function, the RDFs are equal, i.e., \({R}_{S;Z|Y}(\Delta) = {R}_{S;\widehat{X}|Y}(\Delta)\), and are given by the following novel water-filling solution.
\begin{theorem} \label{thm:mainThm}
Suppose that the following conditions hold:\\
(i)  $n = n_x=n_s$ and the inverse of $Q_{X,S|Y}=  Q_{X,S} - Q_{X,Y}Q_Y^{-1}Q_{Y,S}$ exists\footnote{{this can be relaxed using Hotelling's canonical variable form \cite{hotelling1936relations}}},\\
(ii) $Q_{S|Y} \succ 0,\; Q_{X|Y} \succ 0$,\; $Q_{X|Y} \succ Q_{X|S,Y}$. \\
Then define,
\begin{align}
&{\bf Q} \CDC{\tri} Q_{S|Y}^{\frac{1}{2},\CDC{\T}} Q_{X,S|Y}^{-1}=V D U\T,\label{ch1}\\
 &D \tri  \CDC{\mathrm{diag}\big({\bf Q}\big)=}  \mathrm{diag}{\big(d_{1},\dots, d_{n}\big)},\;\; d_{n} \geq d_{n-1} \geq \ldots \geq d_{1},\\
    &Q_{\widehat{X}|Y} \CDC{\tri} Q_{X|Y} - \Sigma =  U \Lambda U\T,\\
    &\Lambda \tri   \mathrm{diag}{\big(\lambda_{1},\dots, \lambda_{n}\big)},\;\;\lambda_{1} \geq \lambda_{2} \geq \ldots \geq \lambda_{n} \geq 0\label{ch2},
\end{align}
where $V$ and $U$ are unitary matrices and $D$ contains the singular values of ${\bf Q}$.
Then, the RDFs are given by 
\begin{align*}
&{R}_{S;Z|Y}(\Delta)= \begin{cases} \infty, &  \Delta = \Delta^- \\  \frac{1}{2}\sum_{i=1}^n\log \Big(\frac{1}{1-\lambda_id_i^2} \Big), & \Delta > \Delta^-  \end{cases} , 
\end{align*}
where $\Delta^- \tri \trace\big(Q_{X|Y}\big) -\sum_{i=1}^n {d_i^{-2}}$,
$\lambda_i$ is given by
\begin{align}
&  \lambda_i  =  {\begin{cases} \frac{1}{d_i^2} - \xi, & \xi \le  \frac{1}{d_i^2}\\ 
 0, & \xi \ge \frac{1}{d_i^2}
 \end{cases}}, \quad i=1,\dots,n
 \label{eq:water},
\end{align}
and  $\xi>0$ is chosen such that $\sum_{i=1}^n \lambda_i = \trace \big(Q_{X|Y}\big) - \Delta$.
\end{theorem}
}
 
\ES{The remainder of the paper is organized as follows: In Section \ref{sec:lower}, we state a lower bound on the conditional mutual information and the sufficient conditions to achieve them. Then, in Section \ref{sec:optireal}, we provide the optimal realization of the RVs $(\widehat{X},Z)$ that achieves the RDFs ${R}_{S;Z|Y}(\Delta)$ and ${R}_{S;\widehat{X}|Y}(\Delta)$. Finally, in Section \ref{sec:water}, we give characterizations of the RDFs via water-filling solutions.}

\section{achievable lower bound on the RDFs} 
\label{sec:lower}
First, we derive an achievable  lower bound on the mutual information $I(S;\widehat{X}|Y)$
for arbitrary RVs.

\begin{theorem} Achievable lower bound on conditional  mutual information \label{them_lb} \\
(a) Let $(X, S, Y, \widehat{X})$ be a quadruple of arbitrary RVs on the abstract spaces  ${\mathbb X}\times {\mathbb S} \times {\mathbb Y}\times \widehat{\mathbb X}$,  with distribution ${\bf P}_{X,S,Y, \widehat{X}}$ and ${\bf P}_{X,S,Y}$  the fixed joint distribution  of $(X, S,Y)$.\\
Define the conditional mean of $X$ conditioned on $(\widehat{X},Y)$ by
$
\overline{X} \tri {\bf E}\big(X\big|\widehat{X},Y\big)$.
Then, the inequality holds,
\begin{equation}
I(S;\widehat{X}|Y) \geq I(S;\overline{X}|Y). \label{eq:LB}
\end{equation}
Moreover if, $\overline{X}=\widehat{X}-a.s$  then \eqref{eq:LB} holds with equality.\\
(b) In part (a) let $(X,S,Y,\widehat{X})$ take values in ${\mathbb R}^{n_x} \times {\mathbb R}^{n_s}   \times {\mathbb R}^{n_y}\times {\mathbb R}^{n_x}$. For all measurable functions, $g(\cdot)$,  $(y, \widehat{x})\longmapsto g(y, \widehat{x})\in {\mathbb R}^{n_x}$ the mean-square error satisfies
\begin{align}
{\bf E}\big (||X-&g(Y, \widehat{X})||^2\big) \geq {\bf E}\big( ||X-\overline{X}||^2\big ),\quad \forall g(\cdot). \label{mse_1}
\end{align}
\end{theorem}\vspace{0.2cm}

\begin{proof}\ES{(a)} The proof is similar to    \cite[Theorem 4]{gkagkosISIT}, with the mutual information $I(X; \widehat{X}|Y)$. (b) This is property of orthogonal projections. 
\end{proof}


Next, we  identify conditions for the RDFs ${R}_{S;\widehat{X}|Y}(\Delta)$ and ${R}_{S;Z|Y}(\Delta)$ to coincide.


\begin{theorem} Equality of the two RDFs\\
\label{rem_lb}
(a) For $Z \in {\pazocal{D}}_{A^\mathrm{c}}(\Delta)$ and $\widehat{X}=f(Y,Z)$, then,
\begin{align}
{R}_{S;Z|Y}(\Delta) \geq R_{S; \widehat{X}|Y}(\Delta).\label{in_1}
\end{align}
(b) In part (a), if the $\widehat{X} \in \pazocal{D}_{A}(\Delta)$, which     achieves $I(S;\widehat{X}|Y)=R_{S; \widehat{X}|Y}(\Delta)$, satisfies $I(S;\widehat{X}|Y)=I(S;Z|Y) $ and $ \widehat{X}=f(Y,Z)$, then,
\bea
 R_{S;Z|Y}(\Delta) = R_{S;\widehat{X}|Y}(\Delta) \label{equality}.
 \eea
 Further,  equality (\ref{equality}) holds  if and only if  any  of the following holds: (i) $I(S;Z|\widehat{X},Y)=0$, (ii)  ${\bf P}_{S,Z|\widehat{X},Y}={\bf P}_{S|\widehat{X},Y}{\bf P}_{Z|\widehat{X},Y}$, 
  (iii) the map   $f(y,\cdot): {\mathbb Z} \rar \widehat{\mathbb X},\;  f(y, z)=\widehat{x}$ is invertible and measurable.
\end{theorem}
\begin{proof} (a) By the chain rule of mutual information,
\begin{align}
I(S;Z|Y)&=I(S;Z,\widehat{X}|Y)  \hso \mbox{by $\widehat{X}=f(Y,Z)$},   \nonumber\\
&=I(S;Z|Y, \widehat{X})+ I(S;\widehat{X}|Y), \nonumber\\
&\geq  I(S;\widehat{X}|Y), \hso \mbox{since $I(S;Z|Y, \widehat{X})\geq 0$.} \label{in_q1}
\end{align}
Since $\widehat{X} \in \pazocal{D}_{A}(\Delta)$  
 the inequality (\ref{in_q1}) is obtained by taking the infimum of over $\pazocal{D}_{A}(\Delta)$. Since $I(S;Z|Y) \geq R_{S; \widehat{X}|Y}(\Delta)$ holds for all $Z \in {\pazocal{D}}_{A^\mathrm{c}}(\Delta)$, by taking the infimum of both sides we obtain~(\ref{in_1}). \\
(b) The first part follows from the fact that,  if $\widehat{X} \in \pazocal{D}_{A}(\Delta)$, which achieves $I(S;\widehat{X}|Y)=R_{S; \widehat{X}|Y}(\Delta)$ is also an element of ${\pazocal{D}}_{A^\mathrm{c}}(\Delta)$, such that $I(S;Z|Y)= I(S;\widehat{X}|Y)$, then necessarily (\ref{equality}) holds. The last part follows from part (a) \CDC{and equivalence of (i)-(iii) from known  properties.} 
\end{proof}

\section{optimal test channel realizations and characterizations of RDFs} \label{sec:optireal}
In this section we consider the following: 
\begin{align}
  (X,S, Y)\in G(0, Q_{(X, S, Y)}), \quad  d(x,\widehat{x})=  ||x-\widehat{x}||^2 \label{eq:gsource}.
\end{align}
\ES{Moreover, we make use of the following properties.}

\begin{proposition}
\label{prop_gaussian_1} 
 Let $X: \Omega \rightarrow {\mathbb R}^{n}$, $n \in {\mathbb Z}_+$,  $X \in G(0, Q_X)$, $Q_{X} \succeq 0$ 
and $S \in {\mathbb R}^{n_1\times n}, n_1 \in {\mathbb Z}_+$. Denote by ${\cal F}^X$ and ${\cal F}^{S X}$ the $\sigma$-algebras generated by the RVs $X$ and $S X$ respectively. The following hold.\\
(a) ${\cal F}^{SX}\subseteq {\cal F}^X$. \\
(b) ${\cal F}^{SX}= {\cal F}^X$ if and only if $\ker(Q_X)=\ker(SQ_X)$.\\
(c) Suppose that $\mathrm{rank}(Q_{X})=n_1, n_1 \in {\mathbb Z}_+,  n_1 <n$. Then, there exists an 
$S \in {\mathbb R}^{n_1\times n}$ such that, if $X_1: \Omega \rightarrow {\mathbb R}^{n_1}$,  $X_1= S X$, then  $X_1 \in G(0; Q_{X_1}), Q_{X_1} \succ 0$ and ${\cal F}^X={\cal F}^{X_1}$.
\end{proposition}
\begin{proof}
 Well-known in measure  theory, see e.g. \cite{SchuppenBook}.
\end{proof}

%

\subsection{Characterization of  RDF with Side Information at Encoder and Decoder}
In this section we characterize the RDF when side information is available both at the encoder and the decoder. First, we derive a {\it preliminary parametrization} of the optimal reproduction distribution ${\bf P}_{\widehat{X}|S, Y}$.

\begin{lemma} \label{lemma:par}
Consider the RDF  ${R}_{S;\widehat{X}|Y}(\Delta)$ defined by (\ref{eq:OP1ck1})
for \eqref{eq:gsource}. Then, the following statements hold.\\
(a) A jointly Gaussian distribution ${\bf P}_{X,S,Y, \widehat{X}}$  minimizes $I(S; \widehat{X}|Y)$, subject to the average distortion.\\
(b)  The conditional reproduction distribution ${\bf P}_{\widehat{X}|S,Y}$ is induced by the  parametric realization of $\widehat{X} \in G(0,Q_{\widehat{X}})$ (in terms of $H, G, Q_W$),
\begin{align}
&\widehat{X} = H S + G Y + W,\hso H \in \mathbb{R}^{n_x\times n_s},\;\;  G \in \mathbb{R}^{n_x\times n_y},  \label{eq:real}\\
&W \in G(0, Q_W), \; Q_W \succeq 0,\; W \; \mbox{indep.   of $(X,S,Y)$. }\label{eq:real_4}
\end{align}
(c)  Consider part (b) and suppose there exist matrices $(H, G, Q_W)$ such that, $\overline{X} = \widehat{X}$-a.s..
 Then, $R_{S;\widehat{X}|Y}(\Delta)$  
 is given by \eqref{eq:OP1ck1} where the infimum is now taken over ${\overline{\pazocal{D}}}_{\mathrm{A}}(\Delta)$ defined by:
\begin{align}
{\overline{\pazocal{D}}}_{\mathrm{A}}(\Delta) \tri \big\{ \widehat{X}: \Omega \rar \widehat{\mathbb X} \big |  \; &(\ref{eq:real}) \;\text{and}\; (\ref{eq:real_4}) \; \mbox{hold},\;\; \overline{X}=\widehat{X}-a.s,\nonumber \\  & {\bf E}\big(||X-\widehat{X}||^2\big)\leq \Delta \big\}.\nonumber 
\end{align}
\end{lemma}
\begin{proof}(a) This  is similar to the classical RDF $R_{X}(\Delta)$ of a Gaussian RV with square-error distortion. (b) By  (a), the test channel distribution ${\bf P}_{\widehat{X}|S,Y}$ is conditionally Gaussian with linear conditional mean and non-random covariance $\Cov(X,\widehat{X}|X)$. Such a distribution is induced by the realizations (\ref{eq:real}), (\ref{eq:real_4}). (c) Due to Theorem \ref{them:lb_g} and (\ref{mse_1}).
\end{proof}


Next we identify sufficient conditions for the lower bounds of Theorem~\ref{them_lb} to be achievable.

\begin{theorem}\label{them:lb_g}
Consider the statement of Theorem~\ref{them_lb} for the  quadruple  of  jointly Gaussian RVs  $(X, S, Y, \widehat{X})$.  
Then, 
\begin{align}
\overline{X}\tri &{\bf E}\big ( X\big|Y, \widehat{X}\big ) =e(Y, \widehat{X}), \label{eq_st_new} \\
=& {\bf E}\big(X\big|Y\big)+\Cov(X,\widehat{X}|Y) \big 
(\Cov(\widehat{X},\widehat{X}|Y)\big )^{\dagger}\big(\widehat{X} - {\bf E}\big(\widehat{X}\big|Y\big)\big). \nonumber  
\end{align}
 Moreover, the following statements hold. \\
Case (i). Suppose  $\Cov(\widehat{X},\widehat{X}|Y)\succ 0$, i.e., $\mathrm{rank}(Q_{\widehat{X}|Y})=n_x$. Then, \eqref{eq:LB} holds with equality if:
\begin{align}
 \overline{X} \tri  {\bf E}\big(X\big|Y, \widehat{X}\big) = e(Y, \widehat{X})=\widehat{X}-a.s..  \label{eq_st}
\end{align} 
The following conditions are sufficient for (\ref{eq_st}) to hold:
\begin{align*}
&\mbox{Condition 1.}\hso 
{\bf E}\big(X\big|Y\big) ={\bf E}\big(\widehat{X}\big|Y\big),\\
&\mbox{Condition 2.}\hso \Cov(X,\widehat{X}|Y) \big (\Cov(\widehat{X},\widehat{X}|Y)\big )^{-1} = I_{n_x} .
\end{align*}
Case (ii). Suppose  $ \Cov(\widehat{X},\widehat{X}|Y)\succeq 0$, i.e., $\mathrm{rank}(Q_{\widehat{X}|Y})=n_1< n_x$. Then, \eqref{eq:LB} holds with equality if:
\begin{align}
\mbox{Condition 3.} \hso& \mbox{For a fixed $y \in {\mathbb Y}$ the function $e(y,\cdot): \widehat{\mathbb X} \rightarrow {\mathbb X}$ } \nonumber \\
&\mbox{$e(y,\widehat{x})=\overline{X}$ uniquely defines $\widehat{x}$.} \label{cond_test_new}
\end{align}
That is,  $e(y,\cdot)$ is  injective  on the support of $\widehat{x}$. A sufficient condition for (\ref{cond_test_new}) to hold is, for a fixed $Y=y \in {\mathbb Y}$, the $\sigma$-algebras satisfy ${\cal F}^{\widehat{X}}= {\cal F}^{e(Y,  \widehat{X})}\big|_{Y=y}$. The function $e(\cdot,\cdot)$ defined in \eqref{eq_st_new} satisfies condition \eqref{cond_test_new}.
\end{theorem} 
\begin{proof} The expression in (\ref{eq_st_new}) follows by properties of jointly Gaussian RVs. Case (i). Clearly,  (\ref{eq_st}) implies $I(S;\widehat{X}|Y) = I(S;\overline{X}|Y)$ due to Theorem~\ref{them_lb}. Case (ii). This is an application of Proposition~\ref{prop_gaussian_1}. 
\end{proof}

Next, we identify the optimal realization, i.e., $(G, H, Q_W)$,  such that   $ \overline{X}=\widehat{X}-a.s$, 
and characterize the RDF. 

\begin{theorem}\label{thm:proof2_ed}
Consider the RDF ${R}_{S;\widehat{X}|Y}(\Delta)$ defined by (\ref{eq:OP1ck1}) for \eqref{eq:gsource}.
Let 
\begin{align*}
    \Sigma\tri  {\bf E} \big ( \big( X - \widehat{X} \big) \big(X - \widehat{X} \big)\T \big ).
\end{align*}
Then, the following statements hold. \\
(a) The  test channel representation  of  $\widehat{X}$ which achieves  ${R}_{S;\widehat{X}|Y}(\Delta)$ is   given by the parametric  realization 
\begin{align}
\widehat{X} &=HS + {\bf E} \big(X\big|Y\big)- H {\bf E} \big(S\big|Y\big) +W, \nonumber \\
 &=HS +  \big( Q_{X,Y} -HQ_{S,Y}  \big)Q_Y^{-1}Y+W, \label{eq:realization} 
 \end{align}
where $H \in {\mathbb R}^{n_s\times n_x}$, $ W \in  G(0, Q_W), \; Q_W \succeq 0$, $W$ is independent of $(X,S,Y)$ and  the parameters $(H, Q_W)$ satisfy,  
 \begin{align}
 &H Q_{X,S|Y}\T = Q_{X|Y} -   \Sigma,  \;\;\; H Q_{X,S|Y}\T= Q_{X,S|Y} H\T \succeq 0, \label{eq:realization_nn_1} \\
&Q_W=  H Q_{X,S|Y}\T-HQ_{S|Y}H\T=Q_{X|Y} -  \Sigma -HQ_{S|Y}H\T, \label{eq:realization_nn_1_a} 
\end{align}
where $Q_{S|Y}\succeq 0$, $Q_{X|Y} \succeq 0$ and $
Q_{X,S|Y} = Q_{X,S} - Q_{X,Y}Q_Y^{-1}Q_{Y,S}$
are fixed matrices.\\
(b) The characterization of the RDF ${R}_{S;\widehat{X}|Y}(\Delta)$ is given by 
\begin{align}
{R}_{S;\widehat{X}|Y}(\Delta) 
&= \inf_{  \substack{\Sigma\succeq 0, \:  \trace ( \Sigma)\leq{\Delta},  \\ Q_{S|Y} \succeq Q_{S|\widehat{X},Y}\succeq 0  }}   \frac{1}{2}\log\Big( \frac{\det\big( Q_{S|Y}\big)}{\det\big( Q_{S|\widehat{X},Y}\big)}\Big),\nonumber
\\
&= \inf_{ \substack{\Sigma\succeq 0,  \:  \trace ( \Sigma)\leq{\Delta},  \\ Q_{\widehat{X}|Y} \succeq Q_W\succeq 0  }}   \frac{1}{2}\log\Big( \frac{\det\big( Q_{\widehat{X}|Y}\big)}{\det\big( Q_W\big)}\Big), \label{rdfchara}
\end{align}
where  
\begin{align*}
&(H,Q_W) \hso \mbox{satisfy  (\ref{eq:realization_nn_1}) and (\ref{eq:realization_nn_1_a})},\\
&Q_{S|\widehat{X},Y} = Q_{S|Y} - Q_{S|Y}H\T Q_{\widehat{X}|Y}^{\dagger} H Q_{S|Y}\succeq 0, \\
&Q_{\widehat{X}|Y}= H Q_{S|Y}H\T +Q_W=Q_{X|Y} -   \Sigma\succeq 0, 
\end{align*}
and it is achieved by the test channel realization of part (a).
\end{theorem}
\begin{proof}  (a) By Lemma~\ref{lemma:par}, and using   (\ref{eq_st_new}) and (\ref{eq_st}), we obtain after some algebra  (\ref{eq:realization_nn_1}) and (\ref{eq:realization_nn_1_a}).  (b) This  follows from the realization. 
\end{proof}
\ES{We deduce the following structural properties.}
\begin{corollary} The test channel specified by Theorem~\ref{thm:proof2_ed} satisfies the following structural properties:
\begin{align*}
&(i) \hso {\bf P}_{X|\widehat{X}, Y}={\bf P}_{X|\widehat{X}},  \hso \mbox{if $Q_{X|Y} \succ \Sigma$},   \\
&(ii) \hso  {\bf P}_{\widehat{X}|S,Y,X}=  {\bf P}_{\widehat{X}| S, Y}, \hso {\bf P}_{X|S,Y,\widehat{X}}=  {\bf P}_{X|S,Y}, \\
&(iii) \hso {\bf E}\big(X\big|\widehat{X}, Y\big)={\bf E}\big(X\big|\widehat{X}\big)=\widehat{X},\hso \mbox{if $Q_{X|Y} \succ \Sigma$}, \\
&(iv) \hso  {\bf E}\big(X\big|Y\big)= {\bf E}\big(\widehat{X}\big|Y\big), \hso \mbox{if $Q_{X|Y} \succ \Sigma$}.
\end{align*} 
\end{corollary}
\begin{proof}
    They can be verified using the realization. 
\end{proof}
\subsection{Characterization of RDF with Side Information \CDC{Only  at the}  Decoder }
In this section we characterize the RDF when side information is available only at the decoder.

\begin{theorem} \label{them_dec} 
{Consider the RDF ${R}_{S;Z|Y}(\Delta)$ defined by (\ref{rdf_d1}) and the RDF ${R}_{S;\widehat{X}|Y}(\Delta)$ characterize in Theorem~\ref{thm:proof2_ed}, both for \eqref{eq:gsource}.}
The following statements hold.  \\
(a) The  test channel representations of  $Z$ and $\widehat{X}=f(Z,Y)$ which achieve  ${R}_{S; Z|Y}(\Delta)$ are,  
\begin{align}
&Z=HS + W, \label{zreal} \\ &\widehat{X}=  \big( Q_{X,Y} -HQ_{S,Y}  \big)Q_Y^{-1}Y + Z, \nonumber
\end{align}
where, $(H, Q_W)$ satisfy the conditions (\ref{eq:realization_nn_1}) and  (\ref{eq:realization_nn_1_a}). \\
(b) The equality ${R}_{S;Z|Y}(\Delta) = {R}_{S;\widehat{X}|Y}(\Delta)$ holds, and it is attained through the test channel realization given in part~(a).
\end{theorem}
\begin{proof}(a) We  re-write \eqref{eq:realization} as \eqref{zreal} by setting
$Z = HS+W$. Therefore, by the above realization $\widehat{X} = f(Z,Y)$ we have \CDC{established} that the inequality \eqref{in_1} holds. Moreover, the inequality in \eqref{in_1} holds with equality, if there exists an $\widehat{X}= f(Z,Y)$  such that   $I(S;Z|\widehat{X},Y)=0$, and  the average  distortion is satisfied.
Taking $\widehat{X}= f_1(Y,Z)$, $Z= f_2(S,W)$ as in \eqref{zreal} then $I(S;Z|\widehat{X},Y)=0$ and the average distortion is satisfied, \CDC{ hence (a) is shown.}   Moreover, item (b) follows from part (a) and  $Q_{Z|Y} = Q_{\widehat{X}|Y}$ and $Q_{S|Z,Y} = Q_{S|\widehat{X},Y}$.  
\end{proof}

\ES{We deduce the following structural properties.}

\begin{corollary} The test channel specified by Theorem~\ref{them_dec} satisfies the following structural properties:
\begin{align*}
&(i) \hso  {\bf P}_{X|\widehat{X},Y,Z}={\bf P}_{X|\widehat{X},Y}={\bf P}_{X|\widehat{X}}, \hso \hso \mbox{if $Q_{X|Y} \succ \Sigma$},\\
& (ii) \hso  {\bf P}_{Z|X, S,Y}=  {\bf P}_{Z| S}, \quad (iii) \hso I(S; Z|Y)= I(S; \widehat{X}|Y),\\
& (iv)\hso Q_{S|Z,Y}= Q_{S|\widehat{X},Y},\;Q_{\widehat{X}|Y}=Q_{Z|Y}  \\
&(v)   \hso {\bf E}\big(X\big|\widehat{X}, Y,Z\big)={\bf E}\big(X\big|\widehat{X}\big)=\widehat{X}, \hso \mbox{if $Q_{X|Y} \succ \Sigma$,}
 \\
&  (vi) \hso   {\bf E}\big(X\big|Y\big)= {\bf E}\big(\widehat{X}\big|Y\big), \hso \mbox{if $Q_{X|Y} \succ \Sigma$}. 
\end{align*}  
\end{corollary}
\begin{proof}
    They can be verified using the realization. 
\end{proof}

Next, we discuss how the RDFs can be used for distributed estimation.

\begin{remark} Distributed estimation  application \label{rem:appl_f}\\
Suppose the noisy channel $C_0$ is Gaussian, and the side information $Y$ is another observation of $X$ through a different noisy Gaussian channel $C_1$:
\begin{align*}
 &S=C_0 X + V_0,\quad  V_0 \in G(0, Q_{V_0}),   \\
 & Y=C_1 X +V_1,\quad  V_1 \in G(0, Q_{V_1}),
\end{align*}
where, $X$, $V_0$ and $V_1$ are {mutually} independent.
We wish to design an estimator of $X$ called $\widehat{X}$ by optimally processing $S$, while $Y$ is available to the estimation strategy that generates $\widehat{X}$, subject to ${\bf E}\{||X-\widehat{X}||^2\} \leq \Delta$. Theorem~\ref{them_dec} states that the optimal processing of $S$ is generated by Gaussian observation RV $Z=HS+W$, which is available to the decoder. Then, the optimal estimation is given by,
\begin{align*}
 \widehat{X}=f(Y, Z)={\bf E}\{X|Y\}- H{\bf E}\{S|Y\}+ Z,   
\end{align*}
where $Z$ is the optimal processing of $S$.
{In other words, rate distortion estimation aims to construct an optimal estimator by constructing the optimal measurement model, i.e., $Z=HS+W$,  which  processes the available information to obtain  the estimator $\widehat{X}$}, contrary to classical Bayesian estimation in which the observations model is given a priori. {In addition, as shown via the  water-filling solution of the RDF in  Section~\ref{sec:water}, RDF estimation ensures the average error is satisfied by \CDC{optimally}  avoiding to estimate  all elements of the vector $X$.}
\end{remark}

Next, we verify that the characterization and realizations of the RDFs degenerate to well know RDFs. 

\begin{remark}
\label{rem_deg}
It is easy to verify from  Theorem~\ref{thm:proof2_ed} and \ref{them_dec} that the realizations and RDFs satisfy the following. \\
(i) If  $ X=S$-a.s. they degenerate to multivariate Wyner RDF \cite{gkagkosISIT} and for scalar RVs to Wyner's RDF \cite{wyner1978}. \\ 
(ii) If $X=S$-a.s. and $Y$ is nonrandom, they degenerate to the classical RDF of IID sources \cite{Cover-2006}. \\
 (iii) For scalar-valued RVs our RDFs degenerate to the RDFs  given by Yamamoto et al. \cite{yamamoto1980source} and Draper and Wornell \cite{draper-wornell2004}.  
\end{remark}\vspace{-0.35cm}
\ES{\begin{remark}It was shown in \cite{gkagkosISIT} and \cite{gkagkos2024structural} that the RDFs and test channel realizations in \cite{tian-chen2009, Zahedi-Ostegraard-2014} are incorrect. Specifically, when \(S = X\) almost surely, the auxiliary variable \(Z\) from \cite{tian-chen2009, Zahedi-Ostegraard-2014} does not match Wyner's formulation \cite{wyner1978, gkagkosISIT}, leading to discrepancies in the RDFs.
\end{remark}}\vspace{-0.1cm}

\section{Characterizations of RDFs via Water-Filling Solutions} \label{sec:water}
In this section, we establish a water-filling solution for the RDFs presented in Theorems \ref{thm:proof2_ed} and \ref{them_dec}. Initially, we derive a lower bound on the RDFs and demonstrate its achievability.
We make use of the following lemma.

\begin{lemma} \label{cor:sp_rep}
Consider the optimal test channel  realizations and characterizations of the RDFs $R_{S;\widehat{X}|Y}(\Delta)$ and $R_{S;Z|Y}(\Delta)$ given in Theorem~\ref{thm:proof2_ed} and \ref{them_dec}, respectively, \ES{ and suppose that (i) and (ii) in Theorem~\ref{thm:mainThm} hold.}
Then,
\begin{align*}
 &H= Q_{\widehat{X}|Y} Q_{X,S|Y}^{-\CDC{\T}},\quad Q_{\widehat{X}|Y} = Q_{X|Y} -\Sigma \succeq 0,\\
&Q_{S|\widehat{X},Y}= Q_{S|Y}   -  Q_{S|Y} H\T Q_{\widehat{X}|Y}^{\dagger}H Q_{S|Y} \succeq 0, 
\end{align*}
and the following hold.\\
 (a) The lower bound holds,
\begin{align*}
 \frac{\det\big (Q_{S|Y}\big )}{\det\big (Q_{S|\widehat{X},Y}\big )} =& \frac{\det\big (Q_{S|Y}\big )}{\det\big (Q_{S|Y}- Q_{S|Y} Q_{X,S|Y}^{-1} \big(Q_{X|Y}-\Sigma\big) Q_{X,S|Y}^{-\CDC{\T}}  Q_{S|Y}\big )} , \nonumber  \\
 \geq & \prod_{i=1}^{n
 } \frac{1}{ \mathrm{diag}\big(I- Q_{S|Y}^{\frac{1}{2},\CDC{\T}} Q_{X,S|Y}^{-1}  Q_{\widehat{X}|Y} Q_{X,S|Y}^{-\CDC{\T}}  Q_{S|Y}^{\frac{1}{2}} \big)}. \label{had_1}
\end{align*}
Moreover, the lower bound above is achieved by the choice of $(H, Q_W)$ such that, \eqref{ch1}-\eqref{ch2} hold. Then, 
\begin{align}
& \frac{\det\big (Q_{S|Y}\big )}{\det\big (Q_{S|\widehat{X},Y}\big )} \geq  \frac{1}{\det\big ( I_{n}-  \Lambda D^2 \big ) }, \hst 0\preceq \Lambda D^2 \preceq I_{n}.
\end{align}
(b) The characterization of the RDF ${R}_{S;Z|Y}(\Delta)$ is given by 
\begin{align*}
{R}_{S;Z|Y}(\Delta)= \inf_{0 \leq  \lambda_i d_i^2 \leq 1, \: \:    \sum_{i=1}^{n} \lambda_i \geq \trace(Q_{X|Y})-\Delta} \frac{1}{2}\sum_{i=1}^{n}\log \Big(\frac{1}{1-\lambda_id_i^2  }\Big).
\end{align*}
\end{lemma}
\ \

\begin{proof}
(a) The lower bound is achieved as follows: Given the matrices $Q_{X|Y}$, $Q_{S|Y}$, and $Q_{X,S|Y}$, one computes the singular value decomposition of ${\bf Q}=V D U\T $. Since $Q_{\widehat{X}|Y}$ is a design parameter, we can choose its eigenvectors to be equal to $U$, that is, $Q_{\widehat{X}|Y} = U\Lambda U\T$. Then, after finding the optimal $\Lambda$, one can compute the distortion matrix as follows: $\Sigma = Q_{X|Y} - U\Lambda U\T$. (b) The characterization follows from item (a) and Theorems~\ref{thm:proof2_ed} and \ref{them_dec}. 
\end{proof}


Finally, we use the characterization of the RDFs given in Lemma~\ref{cor:sp_rep} and compute a their closed-form solutions.
\ES{\begin{proposition} The following statements hold.\\
(a) The RDFs are equal ${R}_{S;Z|Y}(\Delta) ={R}_{S;\widehat{X}|Y}(\Delta) $ and given by Theorem~\ref{thm:mainThm}.\\
(b) The optimal test channel realizations that induce the conditional distributions, which achieve ${R}_{S;\widehat{X}|Y}(\Delta)$ and ${R}_{S;Z|Y}(\Delta)$ of part (a) are given in Theorem \ref{thm:proof2_ed} and \ref{them_dec} respectively.     
\end{proposition}
\begin{proof}
(a) By an application of Karuch-Kuhn-Tucker (KKT) conditions. (b) Follows from Theorem \ref{thm:proof2_ed} and \ref{them_dec}.
\end{proof}
}

\begin{figure}[!t]
\centering
  \centering
  \includegraphics[height = 6cm, width=1\columnwidth]{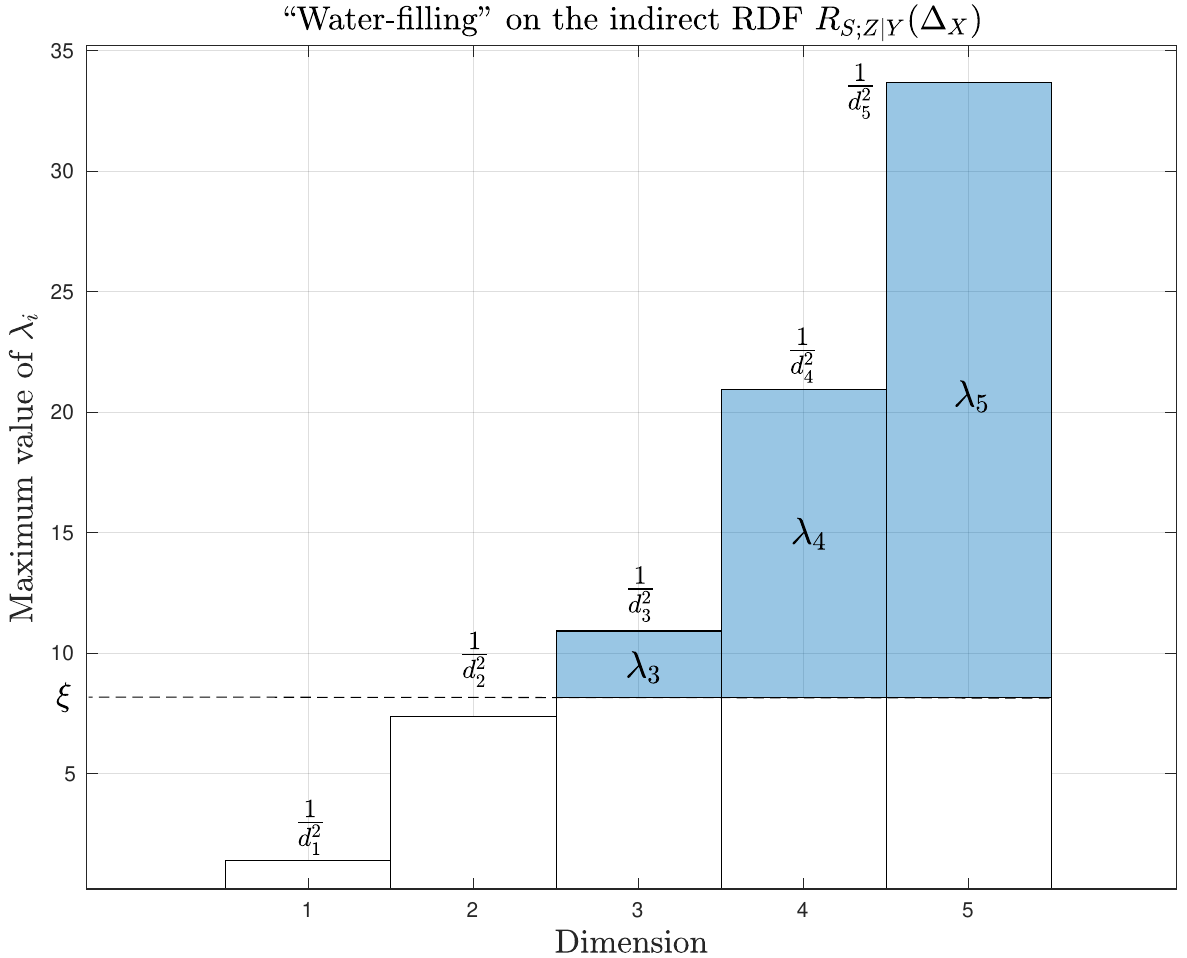}
\caption{Water-filling solution of $\lambda_i,i=1,\dots,5$ for $\Delta = 42$ of Example~1.}  \label{fig:water2} \vspace{-0.55cm}
\end{figure}
\ES{Our water-filling method differs from the classical approach for vector Gaussian sources. Specifically, we compute non-negative elements $\lambda_i$ as $\max(0, 1/d_i^2 - \xi)$, where $d_i$ are the minimizing parameters. In classical water-filling, $1/d_i^2$ is replaced by the variances or eigenvalues of the components. This distinction is due to $\sum_{i=1}^{n} \lambda_i$ representing $\trace(Q_{X|Y}) - \Delta$, not the distortion $\Delta$ itself, which reverses the classical procedure.}

\begin{example}
 { Consider a triple of Gaussian RVs $(X,S,Y)$ such that the matrix $D$ defined in Lemma~\ref{cor:sp_rep} is given by
    \begin{align*}
        D = \diag \big(0.1723, 0.2186, 0.3026,0.3686, 0.8417\big).
    \end{align*}
In this scenario, we find that $\trace(Q_{X|Y}) = 76.1162$. Hence, the maximum distortion value, where the RDFs are zero, is $\Delta_{\max}=76.1162$.} {In Figure \ref{fig:water2}, we depict the water-filling solution of the RDF ${R}_{S;Z|Y}(\Delta)$ as given by \eqref{eq:water} for $\Delta = 42$. The figures indicate that for $i=3,4,5$, we have $\xi > \frac{1}{d_i^2}$, suggesting that $\lambda_i = \frac{1}{d_i^2} - \xi > 0$ for $i=3,4,5$. Consequently, this implies that $\lambda_i = 0$ for $i=1,2$.} {For completeness, in Figure \ref{fig:condRDF},  we plot the values of the  RDFs ${R}_{S;Z|Y}(\Delta)$, as a function of the distortion $\Delta$, which clearly shows a rate that tends to $+\infty$, as $\Delta$ tends to $\Delta^-$.}
\end{example}

\section{Conclusion}
\label{sec:conl}
To summarize, we have derived structural properties of the optimal test channel realizations of the RDFs for the indirect RDF with side information available either at the encoder and decoder or only at the decoder, considering general RVs defined on abstract spaces. Additionally, we have fully characterized the RDFs of jointly Gaussian multivariate RVs with a square-error distortion function using realization theory. Furthermore, we have demonstrated that the two RDFs coincide and have presented a water-filling solution for computing the RDFs. This paper represents the first comprehensive characterization of the two RDFs and provides insights into the structural properties of their optimal test channels.

\begin{figure}[!t]
\centering
  \centering
  \includegraphics[height = 6cm, width=1\columnwidth]{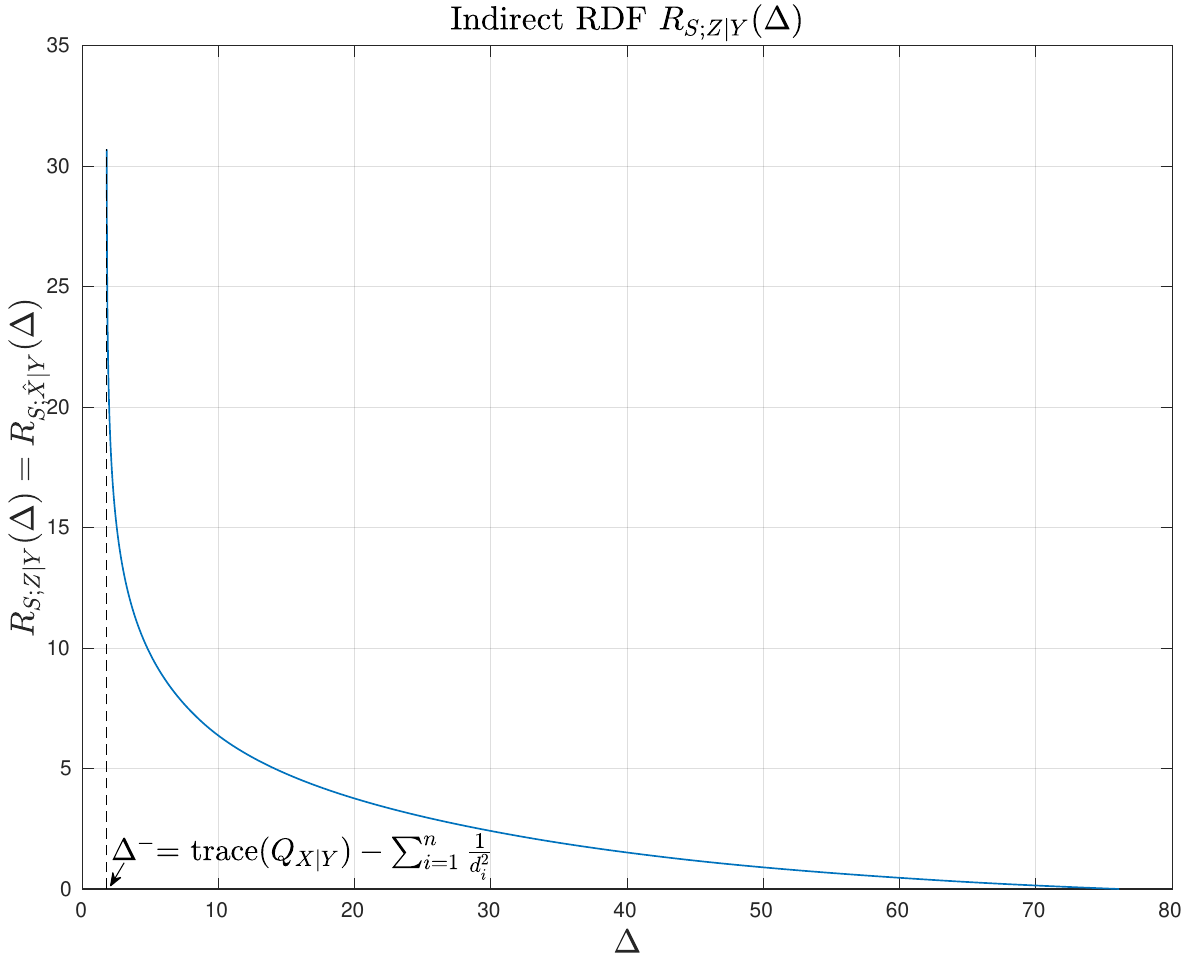}
\caption{The indirect RDF as a function of the distortion $\Delta$ of Example~1.}  \label{fig:condRDF} \vspace{-0.55cm}
\end{figure}


\label{Bibliography}
\bibliographystyle{IEEEtran}
\bibliography{references}

\IEEEtriggeratref{4}



\end{document}